\documentclass[preprint,showkeys,aps]{revtex4}
\usepackage{graphicx}

\begin{document}
\def\be{\begin{equation}}
\def\ee{\end{equation}}
\def\bea{\begin{eqnarray}}
\def\eea{\end{eqnarray}}
\title{MATCHING CONDITIONS IN RELATIVISTIC ASTROPHYSICS}

\author{Hernando Quevedo$^*$}

\address{
Instituto de Ciencias Nucleares, Universidad Nacional Aut\'onoma de M\'exico \\
AP 70543, M\'exico DF 04510, Mexico; \\  
Dipartimento di Fisica, Universit\`a di Roma ``La Sapienza"\\
 Piazzale Aldo Moro 5, I-00185 Roma, Italy;\\
ICRANet, Piazza della Repubblica 10, I-65122 Pescara, Italy\\  
 $^*$E-mail: quevedo@nucleares.unam.mx}
 
\begin{abstract}
We present an exact electrovacuum solution of Einstein-Maxwell equations 
with infinite sets of multipole moments which 
can be used to describe the exterior gravitational field of a rotating charged mass 
distribution. We show that in the special case of a slowly rotating and slightly deformed
body, the exterior solution can be matched to an interior solution belonging to the Hartle-Thorne
family of approximate solutions. To search for exact interior solutions, we propose to use the derivatives of the curvature eigenvalues 
to formulate a $C^3-$matching condition from which the minimum radius can be derived at
which the matching of interior and exterior spacetimes can be carried out. We prove the validity of 
the $C^3-$matching in the particular case of a static mass with a quadrupole moment. The corresponding
interior
solution is obtained numerically and the matching with the exterior solution gives as a result 
the minimum radius of the mass configuration.  

\end{abstract}

\keywords{Matching conditions; Astrophysical Compact Objects.}

\maketitle

\section{Introduction}
\label{sec:int}

In general terms, the group of massive compact objects is defined  in astrophysics to contain black holes, neutron stars, white dwarfs, 
planet-like compact objects,  and other exotic dense stars. The main common characteristic of compact objects is that they are small for their
mass. In the case of weak gravitational fields, Newtonian gravity provides an adequate physical description of compact objects. However,
for the study of very massive objects like black holes and neutron stars it is necessary to take into account relativistic effects. 
Moreover, exotic objects like pulsars and quasars can be investigated only in the framework of relativistic astrophysics. 
Furthermore, the accuracy of modern navigation techniques made it necessary to take into account relativistic effects of the gravitational 
field of the Earth and other planets of the Solar system. 

In general relativity, the gravitational field of compact objects must be described by a metric satisfying Einstein's equations
\be
R_{\mu\nu} - \frac{1}{2} g_{\mu\nu}R = 8\pi T_{\mu\nu} \ ,
\ee
in the interior part of the object ($T_{\mu\nu}\neq 0$) as well as outside in empty space $(T_{\mu\nu}= 0$). Since at the classical level
 the most important sources of gravity are the mass and charge distributions, 
 it is clear that the problem of finding a metric that describes the gravitational 
 field of compact objects 
 can be divided into four related problems. The first one consists in finding an exact vacuum solution $(R_{\mu\nu}=0$) that describes 
the field corresponding to the mass distribution. Secondly, the charge distribution must be considered by solving the electrovacuum 
field equations
\be
R_{\mu\nu}= 8\pi \left( F_{\mu\lambda}F^{\ \lambda}_{\nu } - \frac{1}{4} g_{\mu\nu} F_{\lambda\tau} F^{\lambda\tau}\right)\ ,
\ee
where $F_{\mu\nu}$ is the Faraday tensor corresponding to a charge distribution $Q(x^\mu)$.  
As for the interior region, it
is necessary to propose a model to describe the internal gravitational structure of the object.  
The most popular model is that of a perfect fluid 
\be
T_ {\mu\nu} = (\rho + p )u_\mu u_\nu - g_{\mu\nu} p \ ,
\label{pf}
\ee
with density $\rho(x^\mu)$, pressure $p(x^\mu)$, and 4-velocity $u^\mu$. Finally, the fourth part of the problem consists in considering the 
mass and charge distribution simultaneously, i.e. one needs an exact solution to the equations
\be
R_{\mu\nu} - \frac{1}{2} g_{\mu\nu}R = 8\pi \,  T_{\mu\nu}(\rho, p,Q)\ ,
\ee
where a specific model must be proposed for the energy-momentum tensor. Each of the above problems is very difficult to solve because 
Einstein's field equations are highly non-linear. To overcome in  part these difficulties it is necessary to assume that the gravitational
field is invariant with respect to certain transformations. In particular, one can assume that the field is stationary and axially symmetric, 
conditions that are in good agreement with observations. In fact, compact objects show in general a rotation that 
is constant over quite long periods of time, and their shape deviates only slightly from spherical symmetry.  

In this work, we review a family of electrovacuum solutions that describe the exterior gravitational and electromagnetic fields 
of a rotating compact object. This represents a solution to the first two problems mentioned above. It turns out that in the limit
of a weak gravitational field, a particular solution of this family can be matched smoothly with an interior approximate solution. 
This opens the possibility of searching for an  exact interior solution that could be matched with the exact exterior solution.
As a particular example, we consider the case of a static mass with quadrupole moment, and search for a physically meaningful
interior solution. The problem of matching the solutions arises immediately. It is generally accepted that to match an interior
and an exterior solution one first should determine a matching surface $\Sigma$. If the metrics and their second order derivatives
coincide on  $\Sigma$, the matching is  reached in the sense that the curvature of the interior spacetime region
passes smoothly into the exterior region. This can be considered as a $C^2-$matching. In this procedure, however, the determination of $\Sigma$ 
remains unclear. In the investigation of the example of an interior solution with quadrupole moment reported in this work,
we noticed that at certain distance from the origin of coordinates the curvature eigenvalues present a non-expected behavior in the 
sense that they change their sign several times passing through points of maximum and minimum local curvature. This behavior 
could be associated with the existence of a repulsive gravitational potential. 
It then seems plausible that to avoid 
repulsive gravity one can cover the region of repulsion with an interior solution. The location of $\Sigma$ can then
be anywhere outside the region of repulsion that, in turn, can be invariantly defined in terms of the derivatives of the curvature 
eigenvalues. This can be considered as a $C^3-$matching procedure. In this work, we use this procedure to match the exterior field 
of a mass with quadrupole to the corresponding interior solution. 

\section{The exterior solution}
\label{sec:ext}

Assuming that astrophysical compact objects satisfy the conditions of stationarity and axial symmetry, their gravitational 
and electromagnetic fields can be described by the line element \cite{solutions}
\be
 ds^2 = f (dt-\omega d\varphi)^2 - \frac{\sigma^2}{f}\left[ e^{2\gamma}(x^2-y^2)\left( \frac{dx^2}{x^2-1} + \frac{dy^2}{1-y^2} \right) 
+ (x^2-1)(1-y^2) d\varphi^2\right],
\label{lel}
\ee
where $(t,\varphi,x,y)$ are prolate spheroidal coordinates, and the functions $f$, $\omega$ and $\gamma$ depend on $x$ and $y$ only. 
In general terms, we can say that a solution of the Einstein-Maxwell field equations can be considered as a candidate to describe
the exterior field of a compact object if the metric functions can be written with the functional dependence  
$ f=f(x,y,\tilde P_n)$, $\omega= \omega(x,y,\tilde P_n)$, and $ \gamma = \gamma(x,y,\tilde P_n)$, where $ \tilde P_n = (M_n,J_n,Q_n,H_n), \ n= 0,1,2,...$, represent invariant sets of multipole moments associated with all possible sources of gravitation, namely, $M_n$ are the 
gravitoelectric, $J_n$ the gravitomagnetic, $Q_n$ the electric, and $H_n$ the magnetic moments.  A family of solutions satisfying 
this property was derived in Ref. %
\cite{qm91}. 

It was obtained by using the solution generating techniques which are based upon
the symmetries of the field equations in the Ernst representation \cite{ernst}
\be
(\xi\xi^* - {\cal F} {\cal F}^* -1)\nabla^2 \xi = 2(\xi^*\nabla\xi - {\cal F}^*\nabla {\cal F})\nabla \xi\ ,\ 
\ee
\be
(\xi\xi^* - {\cal F} {\cal F}^* -1)\nabla^2 {\cal F} = 2(\xi^*\nabla\xi - {\cal F}^*\nabla {\cal F})\nabla {\cal F}\ 
\ee
where the gravitational potential $\xi$ and the electromagnetic ${\cal F}$ Ernst potentials are defined as
\be
\xi = \frac{1-f-i\Omega}{1+f+i\Omega}\ , \quad {\cal F}=2\frac{\Phi}{1+f+i\Omega}\ ,
\ee
with   $\sigma (x^2-1)\Omega_x = f^2\omega_y$, and $\sigma (1-y^2)\Omega_y = - f^2 \omega_x$. Moreover, the potential $\Phi$ is determined 
uniquely once the electromagnetic potentials $A_t$ and $A_\varphi$ are given. Finally, $\nabla$ represents the gradient operator in 
prolate spheroidal coordinates. 

The explicit form of the solution can be written as 
\be 
\xi=  \frac{ (a_++ib_+)e^{2\delta\hat\psi} + a_-+i b_- } { (a_+ +ib_+)e^{2\delta\hat\psi} -a_--i b_-}(1-e_0^2 + g_0^2)^{1/2}\ ,
\quad {\Phi}= \frac{e_0+ig_0}{1+\xi}\ , 
\label{epot}
\ee
where
\be 
\hat\psi = \sum_{n=1}^\infty (-1)^n q_n P_n(y)Q_n(x)
\ee
\be
a_\pm = (x\pm1)^{\delta-1} [x(1-\lambda\mu) \pm (1+\lambda\mu)] \ , 
\ee
\be
b_\pm = (x\pm 1)^{\delta -1}[y(\lambda+\mu)\mp(\lambda-\mu)]\ ,
\ee
with
\be
\lambda = \alpha_1 (x^2-1)^{1-\delta}(x+y)^{2\delta -2} e^{2\delta \sum_{n=1}^\infty (-1)^n q_n\beta_{n-}},
\ee
\be
\mu = \alpha_2 (x^2-1)^{1-\delta}(x-y)^{2\delta -2} e^{2\delta \sum_{n=1}^\infty (-1)^n q_n\beta_{n+}}\ ,
\ee
and
\bea
\beta_{n\pm} = & & (\pm 1)^n \left[\frac{1}{2}\ln \frac{(x\mp y)^2}{x^2-1} - Q_1(x)\right] + P_n(y)Q_{n-1}(x)\nonumber \\
               & &-\sum_{k=1}^{n-1} (\pm1)^k P_{n-k}(y)\left[Q_{n-k+1}(x) - Q_{n-k-1}(x)\right]\ .
\eea
Here $P_n(y)$ and $Q_n(x)$ represent the Legendre polynomials and functions of second kind, respectively. The constant
parameters $e_0$, $g_0$, $\sigma$, $\alpha_1$, $\alpha_2$, $q_n$, and $\delta$ determine the gravitational and electromagnetic
multipole moments. The metric functions $f$ and $\omega$ can be obtained from the definitions of the Ernst potentials, whereas
the function $\gamma$ can be calculated by quadratures once $f$ and $\omega$ are known. 
In general, this solution is asymptotically flat and free of singularities along the axis of symmetry, $y=1$, outside 
certain region situated close to the origin of coordinates. The sets of infinite multipole moments can be chosen in such a
way as to reproduce the shape of ordinary axially symmetric compact objects. 

One of the most interesting solutions contained
in this family is the one with non-vanishing parameters $q_0=1$, $q_2=q$, $\delta$, $\alpha_1=\alpha_2= (\sigma - m)/a$, where 
$m$ and $a$ are new constants. Consequently,  the solution possesses the following independent parameters: $m$, $a$, $\delta$, and $q$. 
In the limiting case $\alpha=0$, $a=0$, $q=0$ and $\delta =1$, the only independent parameter is $m$ and the Ernst potential (\ref{epot})
determines the Schwarzschild spacetime. Moreover, for $\alpha =a =0$ and $q=0$ we obtain the Ernst potential of the Zipoy-Voorhees (ZV)
\cite{zip66,voor70}
 static 
solution which is characterized by the parameters $m$ and $\delta$. Furthermore, for   $\alpha =a =0$ and 
$\delta =1$, the resulting solution coincides with the Erez-Rosen (ER) static spacetime \cite{erro59}. 
The Kerr metric is also contained as a special case for $q=0$ and $\delta =1$. 
The physical 
significance of the parameters entering this particular solution 
can be established in an invariant manner by calculating the relativistic Geroch--Hansen \cite{ger,hans} multipole moments.
We use here the procedure formulated in Ref. %
\cite{quev89} which allows us to derive the gravitoelectric $M_n$ as well as the 
gravitomagnetic $J_n$ multipole moments. A lengthly but straightforward calculation yields 
\be 
M_{2k+1} = J_{2k}=0 \ ,  \quad k = 0,1,2,... 
\ee 
\be 
M_0= m + \sigma(\delta -1)
\ee
\be
M_2 = \frac{2}{15} \sigma^3 \delta q - \frac{1}{3}\sigma^3 (\delta^3 -3\delta^2-4\delta + 6) - m\sigma^2 \delta (\delta -2) - 3m^2 \sigma (\delta -1)
-m^3 \ ,
\ee
\be
J_1 = ma + 2a \sigma (\delta -1)\ ,
\ee
\bea
J_3 && = \frac{4}{15} a\sigma^3 \delta q \nonumber \\
&& - a\left[ \frac{2}{3} \sigma^3 (\delta^3 - 3\delta^2 - \delta + 3) + m\sigma^2 (3\delta^2 - 6 \delta + 2)
+ 4 m^2 \sigma (\delta -1) + m^3\right]   .
\eea
The even gravitomagnetic and the odd gravitoelectric multipoles vanish identically because the solution possesses and additional reflection 
symmetry with 
respect to the hyperplane $y=0$ which can be interpreted as the equatorial plane.  
Higher odd gravitomagnetic and even gravitoelectric multipoles can be shown to be linearly dependent since they are completely determined in terms 
of the parameters $m$, $a$, $q$ and $\delta$. From the above expressions we see that the ZV parameter $\delta$  
enters explicitly the value
of the total mass $M_0$ as well as the angular momentum $J_1$ of the source. The mass quadrupole $M_2$ can be interpreted as a nonlinear superposition
of the quadrupoles corresponding to the ZV, ER and Kerr spacetimes. A generalization of the Kerr metric which includes an arbitrary quadrupole
moment is obtained by imposing the condition $\delta=1$. The resulting multipoles are
\begin{equation} 
M_{2k+1} = J_{2k}=0 \ ,  \quad k = 0,1,2,... 
\end{equation} 
\begin{equation} 
M_0 = m \ , \quad M_2 = - ma^2 + \frac{2}{15}qm^3 \left(1-\frac{a^2}{m^2}\right)^{3/2}  \ , ... 
\end{equation} 
\begin{equation} 
J_1= ma \ , \quad J_3 = -ma^3 +  \frac{4}{15}qm^3 a \left(1-\frac{a^2}{m^2}\right)^{3/2}  \ , ....
\end{equation}
It is interesting to note that this particular exact solution in the limit $a\rightarrow m$ leads to the spacetime of an extreme Kerr black hole,
regardless of the value of the quadrupole parameter $q$.

In the limiting static case of the ZV metric the only non-vanishing parameters are $m=\sigma$ and $\delta$ so that all gravitomagnetic 
multipoles vanish and we obtain 
\be 
 M_0= m\delta\ ,\quad 
 M_2 = \frac{1}{3}\delta m^3 (1-\delta^2) \ ,
 \ee
for the leading gravitoelectric multipoles. Consequently, the 
ZV solution represents the exterior gravitational field of a static deformed body. 

In all the above special solutions the electromagnetic field vanishes identically. One can easily obtain the corresponding electrovacuum 
generalizations by assuming that $e_0\neq 0$ and $g_0\neq 0$. The computation of the respective electromagnetic multipole moments can be
performed in an invariant manner and the result can be expressed as
\be
E_n = e_0 M_n \ , \quad H_n = g_0 J_n \ .
\ee
This means that the charge distribution resembles the mass distribution. The electric moments vanish identically if no mass distribution 
exists. This result is in accordance with our physical intuitive interpretation of a charge distribution. The magnetic moments turn out
to be proportional to the gravitomagnetic multipoles, with no magnetic monopole. This is a physical reasonable result in the sense that the 
magnetic field is generated by the motion of the charge distribution, in the present case, by the rotation of the compact object. 

It is worth noticing that in all the above special solutions 
we assumed that $\alpha_1=\alpha_2$ and obtained generalizations of the Kerr metric with arbitrary 
quadrupole moment. More general solutions can be obtained by relaxing this condition. Consider, for instance, the special solution
with $\delta=1$, $q_0=1$, $q_i=0$ for $i=1,2,...$, and
\be
\alpha_1 = \frac{\sigma-\frac{m}{\eta}}{a+\frac{l}{\eta}}\ ,\quad
\alpha_2 = \frac{\sigma-\frac{m}{\eta}}{a-\frac{l}{\eta}}\ ,\quad
\sigma^2 = \frac{m^2+l^2}{\eta^2} - a^2\ , \quad \eta = \frac{1}{\sqrt{1-e_0^2}} \ .
\ee
The resulting potential corresponds to  the charged Kerr-Taub-NUT spacetime with total charge $Q_0 = me_0$, where 
$l$ is the Taub-NUT parameter.  
  
\section{Matching with an approximate interior  solution}
\label{sec:intht}

In general relativity, the gravitational field of a compact object is described by a Riemannian manifold 
which must be well defined in the entire spacetime. Usually, the exterior metrics are characterized by 
the presence of curvature singularities in a very specific region of spacetime. A possibility to eliminate those singularities 
is to find a physically meaningful regular interior solution that covers the singular region. If the two solutions 
can me matched smoothly on some hypersurface $\Sigma$ , one can say that  the entire manifold is well defined. 
In this section we will see that a particular solution contained in the above family determines  a well-defined
manifold in the entire spacetime of a particular rotating compact object.

Consider the solution (\ref{epot}) with the special choice
\be
e_0=g_0=0\ ,\quad q_0 =1 \ , \quad  q_1 =0\ ,\quad  q_2 = q\ ,\quad \delta = 1-q\ ,
\ee
i. e., we consider a vacuum solution with a ZV parameter given in terms of the quadrupole parameter. Moreover, the 
angular momentum parameters are chosen as 
\be
\alpha_1=\alpha_2 =\frac{\sigma-m}{a} \ ,\qquad \sigma^2 = m^2 - a^2\ .
\ee
Furthermore, let us assume that the condition for a slowly rotating slightly deformed compact object is satisfied, i. e., 
\be
|q_j| << |q|\ , \ j=3,4,\ldots \ , \qquad \frac{a^2}{m^2}<< 1 \ .
\ee
Expanding the resulting potential up to the first order in $q$ and up to the second order in $a/m$, and introducing spherical-like
coordinates $R=R(x,y)$ and $\Theta=\Theta(x,y)$, a stationary axisymmetric spacetime is obtained whose line element can be
written as \cite{bglq09,quev10} 
\bea
\label{ht1}
ds^2&=&\left(1-\frac{2{\mathcal M }}{R}\right)\left[1+2k_1P_2(\cos\Theta)+2\left(1-\frac{2{\mathcal M}}{R}\right)^{-1}\frac{J^{2}}{R^{4}}(2\cos^2\Theta-1)\right]dt^2\nonumber \\
&&-\left(1-\frac{2{\mathcal M}}{R}\right)^{-1}\left[1-2k_2P_2(\cos\Theta)-2\left(1-\frac{2{\mathcal M}}{R}\right)^{-1}\frac{J^{2}}{R^4}\right]dR^2\nonumber \\
&&-R^2[1-2k_3P_2(\cos\Theta)](d\Theta^2+\sin^2\Theta d\varphi^2)+4\frac{J}{R}\sin^2\Theta dt d\varphi\ ,
\eea
where
\begin{eqnarray}\label{ht2}
k_1&=&\frac{J^{2}}{{\mathcal M}R^3}\left(1+\frac{{\mathcal M}}{R}\right)+\frac{5}{8}\frac{Q-J^{2}/{\mathcal M}}{{\mathcal M}^3}Q_2^2\left(\frac{R}{{\mathcal M}}-1\right)\ , \nonumber\\
k_2&=&k_1-\frac{6J^{2}}{R^4}\ , \nonumber\\
k_3&=&k_1+\frac{J^{2}}{R^4}-\frac{5}{4}\frac{Q-J^{2}/{\mathcal M}}{{\mathcal M}^2R}\left(1-\frac{2{\mathcal M}}{R}\right)^{-1/2}Q_2^1\left(\frac{R}{\mathcal M}-1\right)\ .\nonumber
\end{eqnarray}
Here $Q_l^m$ are the associated Legendre functions of the second kind
\begin{equation}
Q_{2}^{1}(x)=(x^{2}-1)^{1/2}\left[\frac{3x^{2}-2}{x^{2}-1}-\frac{3}{2}x\ln\frac{x+1}{x-1}\right],
\nonumber
\ee
\be
\ \ Q_{2}^{2}(x)=\frac{3}{2}(x^{2}-1)\ln\frac{x+1}{x-1}+\frac{5x-3x^{3}}{x^{2}-1}.
\nonumber
\end{equation}
The metric (\ref{ht1}) represents the exterior gravitational field of a slowly and rigidly rotating compact object in which the 
deviation from spherical symmetry has been taken into account only up to the first order in the quadrupole parameter $q$. The
main gravitational multipoles are the total mass ${\mathcal M}$, the total angular momentum $J$ and the quadrupole moment $Q$ which
reduce to 
\begin{equation}
{\mathcal M}=m(1-q),\ \ J=-ma,\ \ Q=\frac{J^{2}}{m}-\frac{4}{5}m^{3}q\ .
\end{equation}

For the internal structure of the compact object one can assume a perfect fluid model, 
satisfying a one-parameter equation of state, $p=p(\rho)$. The fact that we are limiting ourselves to the case of slow rotation 
simplifies the problem. In fact, one can first solve the non-rotating case and then 
linearize the field equations  with respect to the angular velocity $\Omega$ which is assumed to be uniform. The 4-velocity $u^\mu$ can
be identified with the velocity of an observer moving with the particles of the fluid so that 
$u^R=u^\Theta=0,$ and $\ u^\varphi=\Omega u^t$. 
The explicit integration of  Einstein's equations depends on the particular choice of the density function $\rho(R)$,
and on the equation of state. For the sake of simplicity, we consider 
here only the simplest case with $\rho=$const so that the pressure $p(R)$ must be determined by the field equations. 
Then, in the limiting case of a non-rotating object the total mass is ${\cal M}=4\pi  \rho {\cal R}^3 / 3$, 
where ${\cal R}$ is the radius of the body. Under these
conditions, the resulting line element can be written as 
\bea
\label{h1}
ds^2 & = &\left(1+{2\Phi}{}\right)dt^2-
\left[1+{2R}{}\frac{d\Phi_{0}(R)}{dR}+\Phi_{2}(R) P_2(\cos\Theta)\right]dR^2\nonumber\\
 && -R^2\left[1+{2\Phi_{2}(R)}{}P_2(\cos\Theta)\right][d\Theta^2+\sin^2\Theta (d\varphi-\tilde{\omega} dt)^2],
\eea
where
\begin{equation}
\Phi=\Phi_{0}(R)+\Phi_{2}(R)P_2(\cos\Theta),
\end{equation}
is the interior Newtonian potential. Here $\Phi_{0}$ is the interior
 Newtonian potential for the  non-rotating configuration and
$\Phi_{2}(R)$ is the perturbation due to the rotation. Moreover, $ \tilde{\omega}=\Omega-\bar{\omega}$
is the angular velocity of the local
inertial frame, where $\Omega=$const is the angular velocity of the fluid 
and $\bar{\omega}=\bar\omega(R)$ is the angular
velocity of the fluid relative to the inertial frame.

It can be shown that the interior solution (\ref{h1}) can be matched smoothly on the surface $R=\mathcal{R}$
with the exterior solution (\ref{ht1}) in the special 
case of an unperturbed configuration with $\Phi_2 =0$ and 
\begin{equation}
\label{phi0}
\Phi_{0}(R)=-2\pi \rho\left({\cal R}^{2}-\frac{R^{2}}{3}\right),
\end{equation}
where $\cal R$ is the radius of the non-rotating configuration.
In the general case 
of a rotating deformed body, the matching can be performed only numerically. In fact, the solutions of the field equations 
for the inner distribution of mass are calculated using the matching conditions as boundary conditions. In this manner, it is
possible to find realistic solutions that describe the interior and exterior gravitational field of slowly rotating
and slightly deformed mass distributions. Several examples of this procedure were originally presented by
Hartle and Thorne \cite{hartle1,hartle2}. 

The above result shows that in the case of a slowly and rigidly rotating compact object that slightly deviates from spherical symmetry, 
the solution presented here is  physically meaningful and determines a well-defined manifold in the entire spacetime. 
We interpret this result as a strong indication that in general the solution (\ref{epot}) describes the exterior gravitational 
field of rotating astrophysical compact bodies with arbitrary 
sets of gravitational and electromagnetic multipole moments. 

\section{Matching with an exact interior  solution}
\label{sec:intzv}

Rather few exact stationary solutions that involve a matter distribution in rotation are to be found in the literature. 
In particular, the interior solution for the rotating Kerr solution is still unknown. In fact, the quest for a realistic exact 
solution, representing both the interior and exterior gravitational field generated by a self-gravitating axisymmetric distribution 
of a perfect fluid mass in stationary rotation is considered as a major problem in general relativity. We believe that the inclusion 
of a quadrupole in the exterior and in the interior solutions adds a new physical degree of freedom that could be used to search for realistic
interior solutions. We will study in this section the entire Riemannian manifold corresponding to the simple case of a static exterior solution 
with only quadrupole moment. 

The simplest generalization of the Schwarzschild spacetime which includes a quadrupole parameter can be obtained from the Zipoy--Voorhees
solution with $\delta = 1-q$. The corresponding line element in spherical-like coordinates can be represented as
\bea
ds^2 = & & \left(1-\frac{2m}{r}\right)^{1-q} dt^2   \\
& -& \left(1-\frac{2m}{r}\right)^{q}\left[ \left(1+\frac{m^2\sin^2\theta}{r^2-2mr}\right)^{q(2-q)} \left(\frac{dr^2}{1-\frac{2m}{r}}+ r^2d\theta^2\right) + r^2 \sin^2\theta d\varphi^2\right]\nonumber .
\label{zv}
\eea

This solution is axially symmetric and reduces to the spherically symmetric Schwarzschild metric in the limit $q\rightarrow 0$. 
It is asymptotically flat for any finite values of the parameters $m$ and $q$. Moreover, in the limiting case $m\rightarrow 0$
it can be shown that the metric is flat. This means that, independently of the value of $q$, there exists a coordinate transformation that 
transforms the resulting metric into the Minkowski solution. From a physical point of view this is an important  
property because it means that the parameter $q$ is related to a genuine mass distribution, i.e., there is no quadrupole moment 
without mass.  
To see this explicitly,
we calculate the multipole moments of the solution by using the invariant definition proposed by Geroch \cite{ger}. The 
lowest mass multipole moments $M_n$, $n=0,1,\ldots $ are given by
\be 
M_0= (1-q)m\ , \quad M_2 = \frac{m^3}{3}q(1-q)(2-q)\ ,
\ee
whereas higher moments are proportional to $mq$ and can be 
completely rewritten in terms of $M_0$ and $M_2$. 
This means that the arbitrary parameters $m$ and $q$ determine the mass and quadrupole 
which are the only independent multipole moments of the solution. In the limiting case $q=0$ only the monopole $M_0=m$ 
survives, as in the Schwarzschild spacetime. In the limit $m=0$, with $q\neq 0$, all moments vanish identically, implying that 
no mass distribution is present and the spacetime must be flat. This is in accordance with the result mentioned above for the 
metric (\ref{zv}). Furthermore, notice that all odd multipole moments are zero because the solution possesses an additional 
reflection symmetry with respect to the equatorial plane. 

We conclude that the above metric describes the exterior gravitational 
field of a static deformed mass. The deformation is described by the quadrupole moment $M_2$ which is positive for a prolate mass 
distribution and negative for an oblate one. Notice that in order to avoid the appearance of
a negative total mass $M_0$ the condition $q<1$ must be satisfied . 

\subsection{Matching conditions}
\label{sec:mat}

In this subsection we analyze several approaches which could be used to determine the matching hypersurface $\Sigma$. 
Instead of presenting a rigorous analysis, we will present an intuitive method based on 
the behavior of the curvature and the motion of test particles.  

\begin{figure}
\includegraphics[scale=0.2]{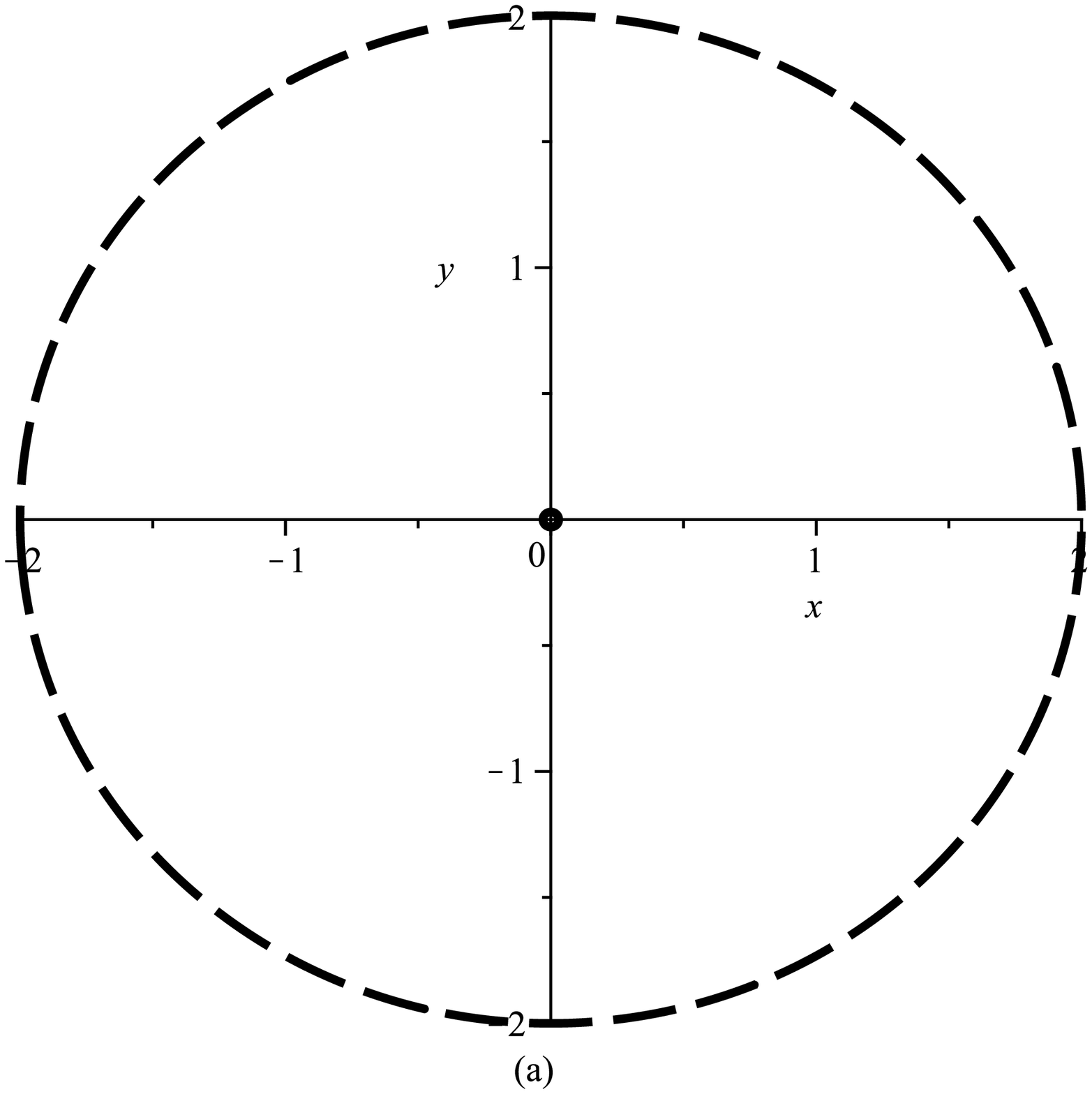}
\includegraphics[scale=0.2]{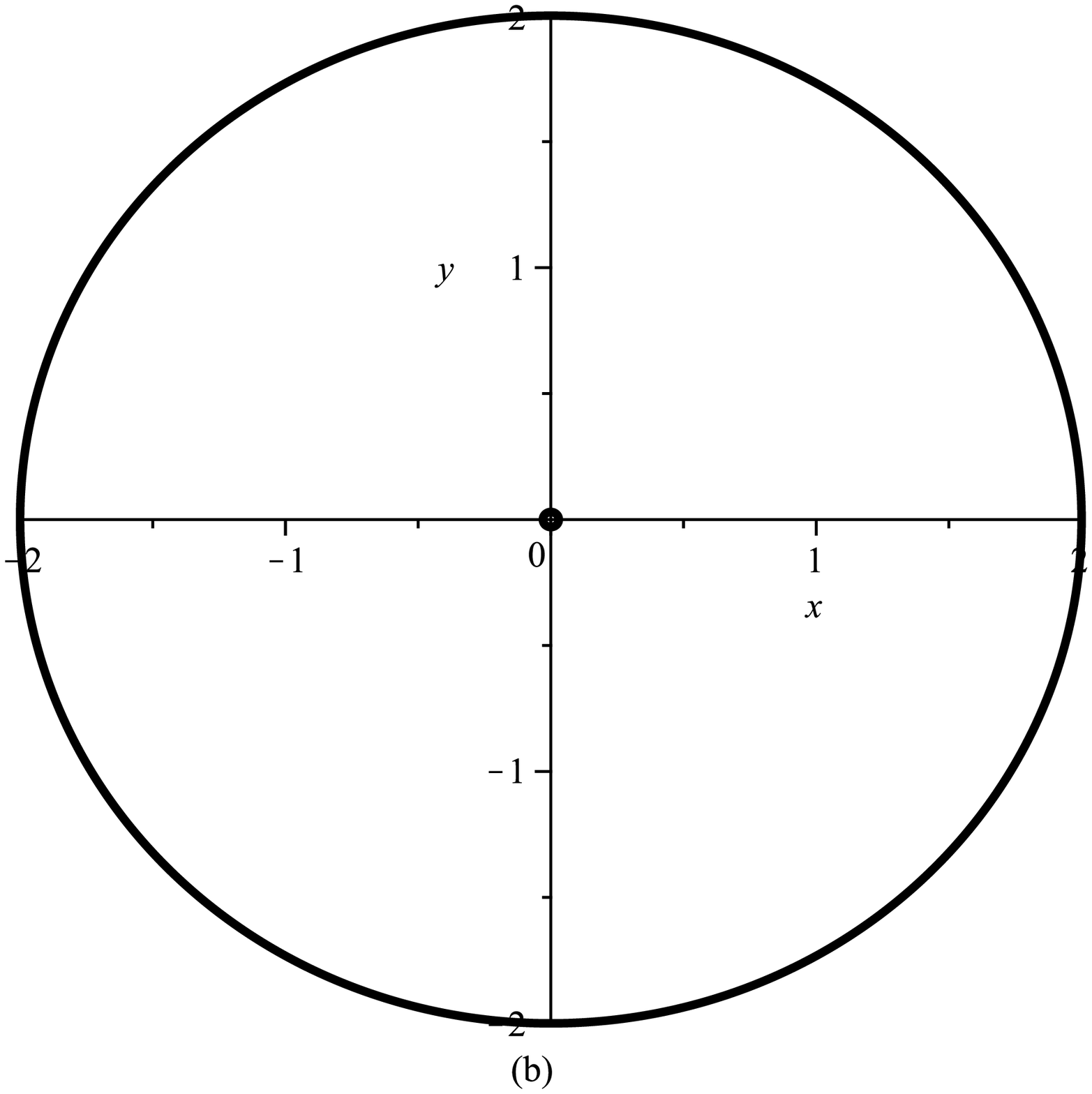}
\includegraphics[scale=0.2]{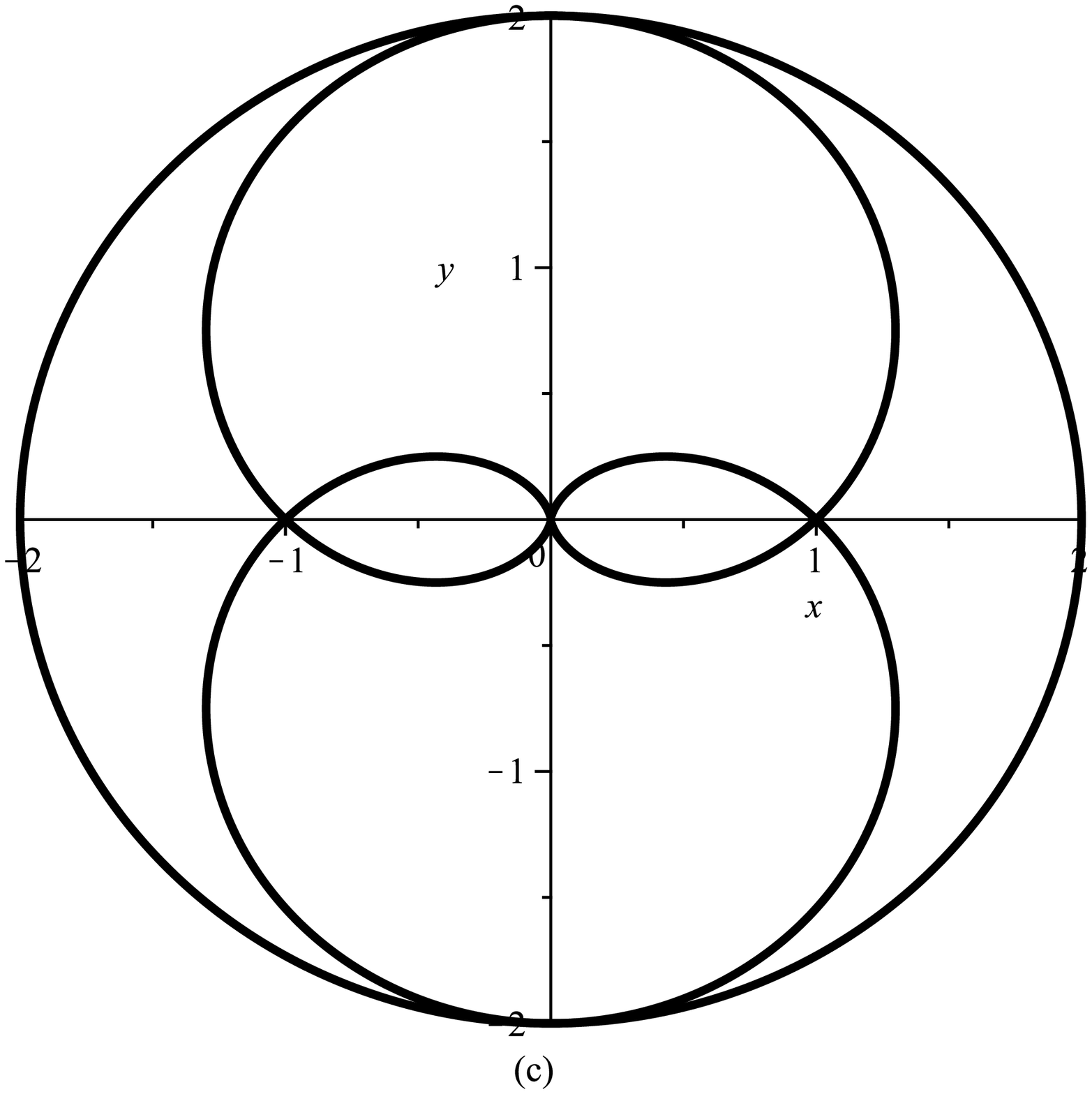}
\caption{Structure of naked singularities of a spacetime with quadrupole parameter $q$. 
Plot (a) represents the limiting case 
of a Schwarzschild spacetime $(q=0)$ with a singularity at the origin of coordinates surrounded by the horizon (dashed curve) 
situated at $r=2m$. Once the quadrupole parameter $q$ is included, the horizon transforms into 
a naked singularity (solid curve) and the central singularity becomes naked as well. This case is illustrated in plot (b). 
For values of the quadrupole parameter within the interval $q\in \left(1- \sqrt{3/2},1+\sqrt{3/2}\right)\backslash \{0\}$, 
two additional
naked singularities appear as depicted in plot (c).}
\label{fig1}
\end{figure}

To investigate the structure of possible curvature singularities, we consider the Kretschmann scalar 
$K = R_{\mu\nu\lambda\tau}R^{\mu\nu\lambda\tau}$. A straightforward computation leads to 
\be
K   = \frac{16 m^2(1-q)^2}{r^{4(2-2q+q^2)}}\frac{ (r^2-2mr+m^2\sin^2\theta)^{2q^2 -4q -1}}{(1-2m/r)^{2(q^2-q+1)}}L(r,\theta)\ ,
\label{kre}
\ee
with 
\bea
L(r,\theta)= & & 3(r-2m+qm)^2(r^2-2mr+m^2\sin^2\theta) \nonumber\\
& & - q(2-q)\sin^2\theta[ q^2 - 2q  + 3(r-m)(r-2m+qm)] \ .
\eea
In the limiting case $q=0$, we obtain the Schwarzschild value $K= {48 m^2}/{r^6}$ with the only singularity 
situated at the origin of coordinates $r\rightarrow 0$. In general, one can show that the singularity at
the origin, $r=0$, is present for any values of $q$. Moreover, an additional singularity appears at the radius $r=2m$ 
which, according to the metric (\ref{zv}), is also a horizon in the sense that the norm of the timelike Killing 
tensor vanishes at that radius. Outside the hypersurface $r=2m$ no additional horizon exists, indicating 
that the singularities situated at the origin and at $r=2m$ are naked. Moreover, for values of the quadrupole parameter 
within the interval
\be 
q\in \left(1- \sqrt{3/2},1+\sqrt{3/2}\right)\backslash \{0\}
\ee
a singular hypersurface appears at a distance 
\be
r_\pm = m(1\pm \cos\theta)
\ee
from the origin of coordinates. This type of singularity is always contained within the naked singularity situated at the radius
$r=2m$, and is related to a negative total mass $M_0$ for $q>1$. Nevertheless, in the interval 
$q\in (1- \sqrt{3/2},1] \backslash \{0\}$ the singularity is generated by a more realistic source with positive mass. 
This configuration of naked singularities is schematically illustrated 
in Fig. \ref{fig1}. 

The analysis of singularities is important to determine the matching hypersurface $\Sigma$. Indeed, in the case under 
consideration it is clear that $\Sigma$ cannot be situated inside the sphere defined by the radius $r=2m$. To eliminate 
all the singularities it is necessary to match the above solution (\ref{zv}) with an interior solution which covers completely 
the naked hypersurface $r=2m$. 

Another important aspect related to the presence of naked singularities is the problem of repulsive gravity.  
In fact, it now seems to be established that naked singularities can appear as the result of 
a realistic gravitational collapse \cite{joshi07} and that naked singularities can 
generate repulsive gravity. Currently, there is no invariant definition of
repulsive gravity in the context of general relativity, although some attempts have been made by using 
invariant quantities constructed with the curvature of spacetime \cite{def89,def08,christian}. Nevertheless, it is possible
to consider an intuitive approach by using the fact that the motion of test particles in stationary axisymmetric
gravitational fields reduces to the motion in an effective potential. This is a consequence of the fact that
the geodesic equations possess two first integrals associated with stationarity and axial symmetry. The explicit 
form of the effective potential depends also on the type of motion under consideration. 

\begin{figure}
\includegraphics[scale=0.3]{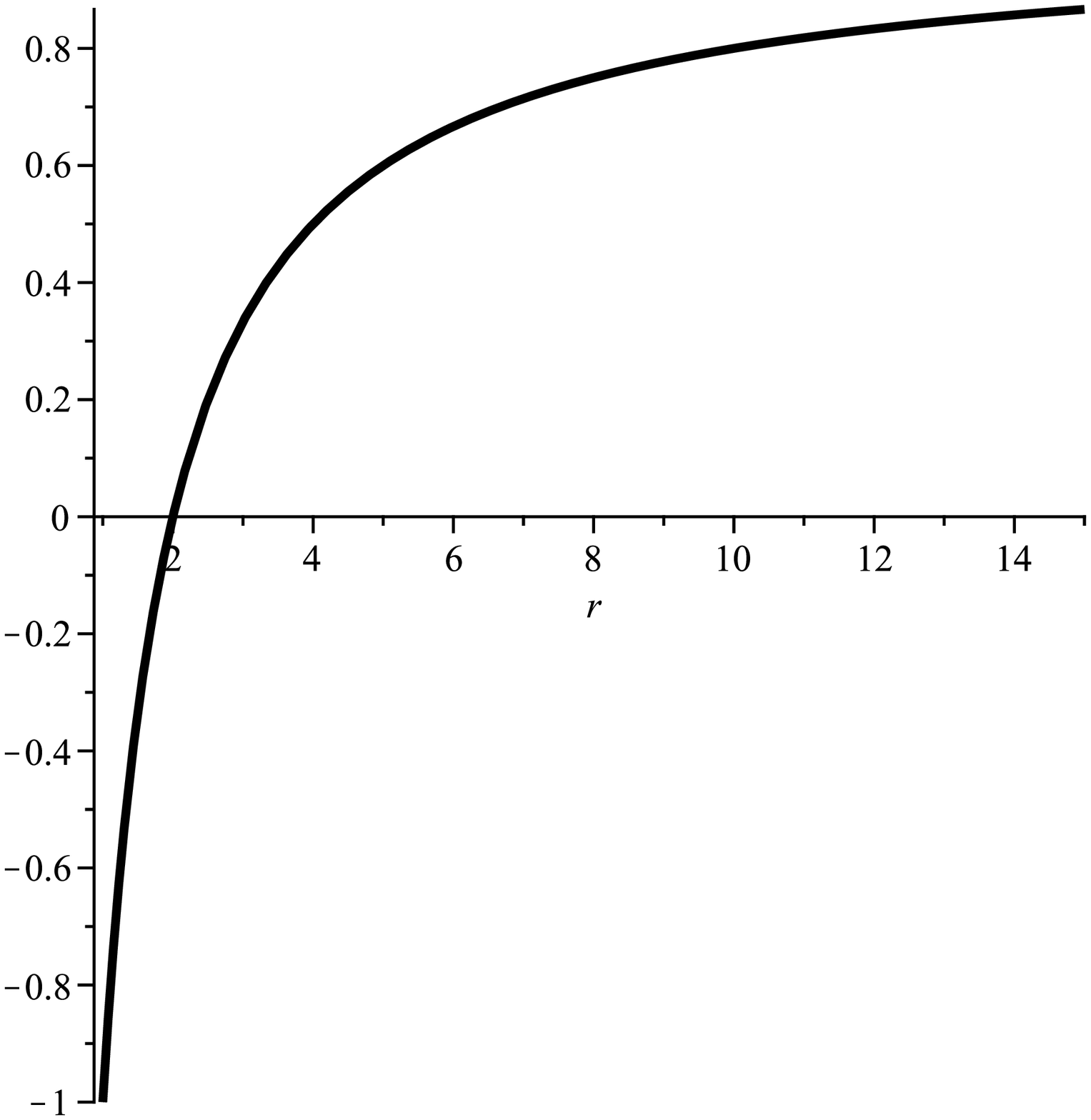}
\includegraphics[scale=0.3]{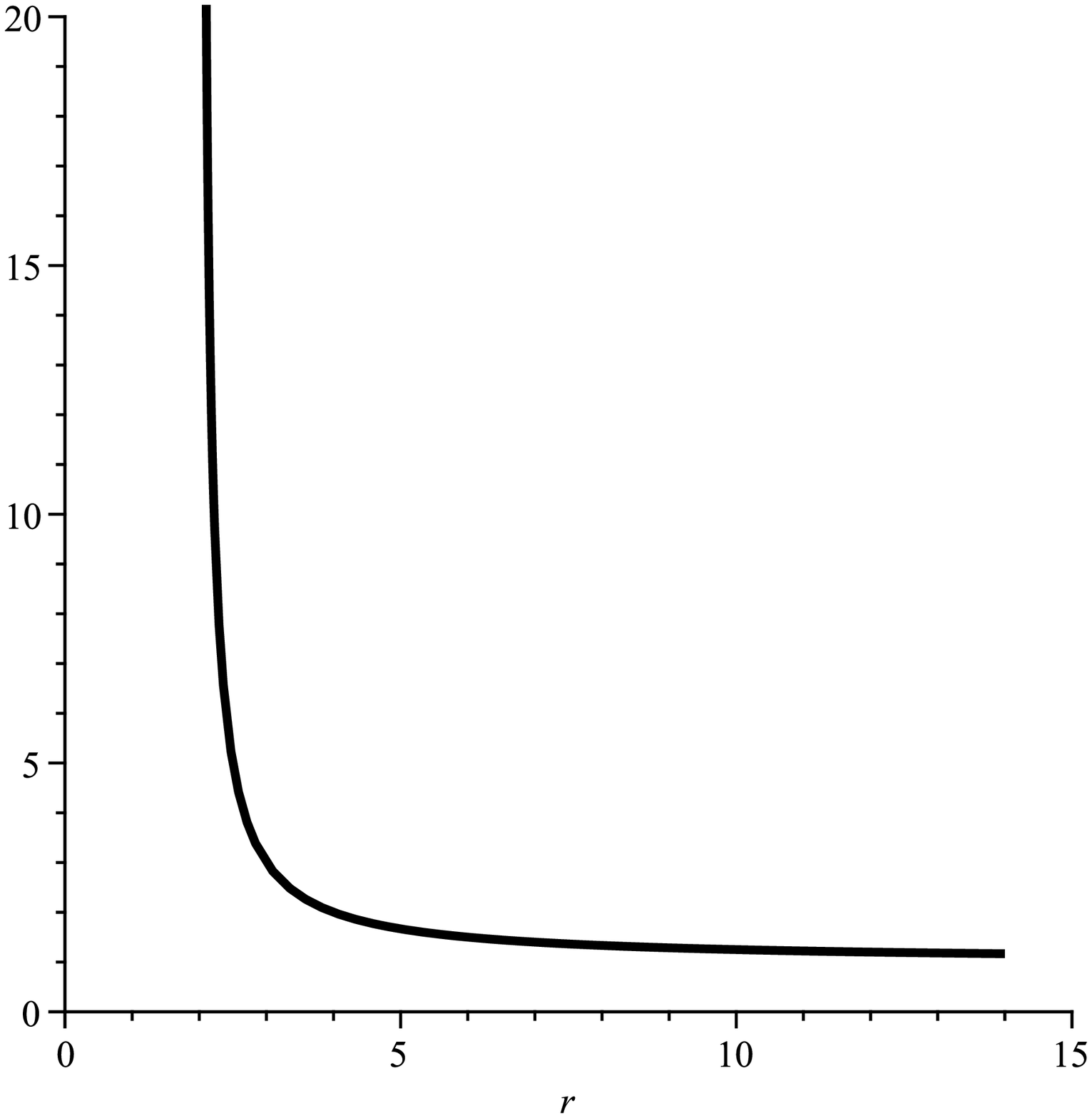}
\caption{The effective potential for the motion of timelike particles. Plot (a) shows the typical behavior of 
the effective potential  of a black hole configuration with $q=0$. The case of a naked singularity with $q=1/2$ 
is depicted in plot (b). } 
\label{fig2}
\end{figure}

In the case of a massive test 
particle moving  along a geodesic contained in the equatorial plane  $(\theta = \pi/2)$ of the Zipoy--Voorhees 
spacetime (\ref{zv}), one can show that the effective potential reduces to
\be
V_{eff}^2 = \left(1-\frac{2m}{r}\right)^{1-q}\left[ 1 + \frac{L^2}{r^2}\left(1-\frac{2m}{r}\right)^{-q}\right]\ ,
\ee
where $L$ is constant associated to the angular momentum of the test particle as measured by a static observer at rest 
at infinity. This expression shows that the behavior of the effective potential strongly depends on the value of the quadrupole 
parameter $q$. This behavior is illustrated in Fig. \ref{fig2}. 

\begin{figure}
\includegraphics[scale=0.3]{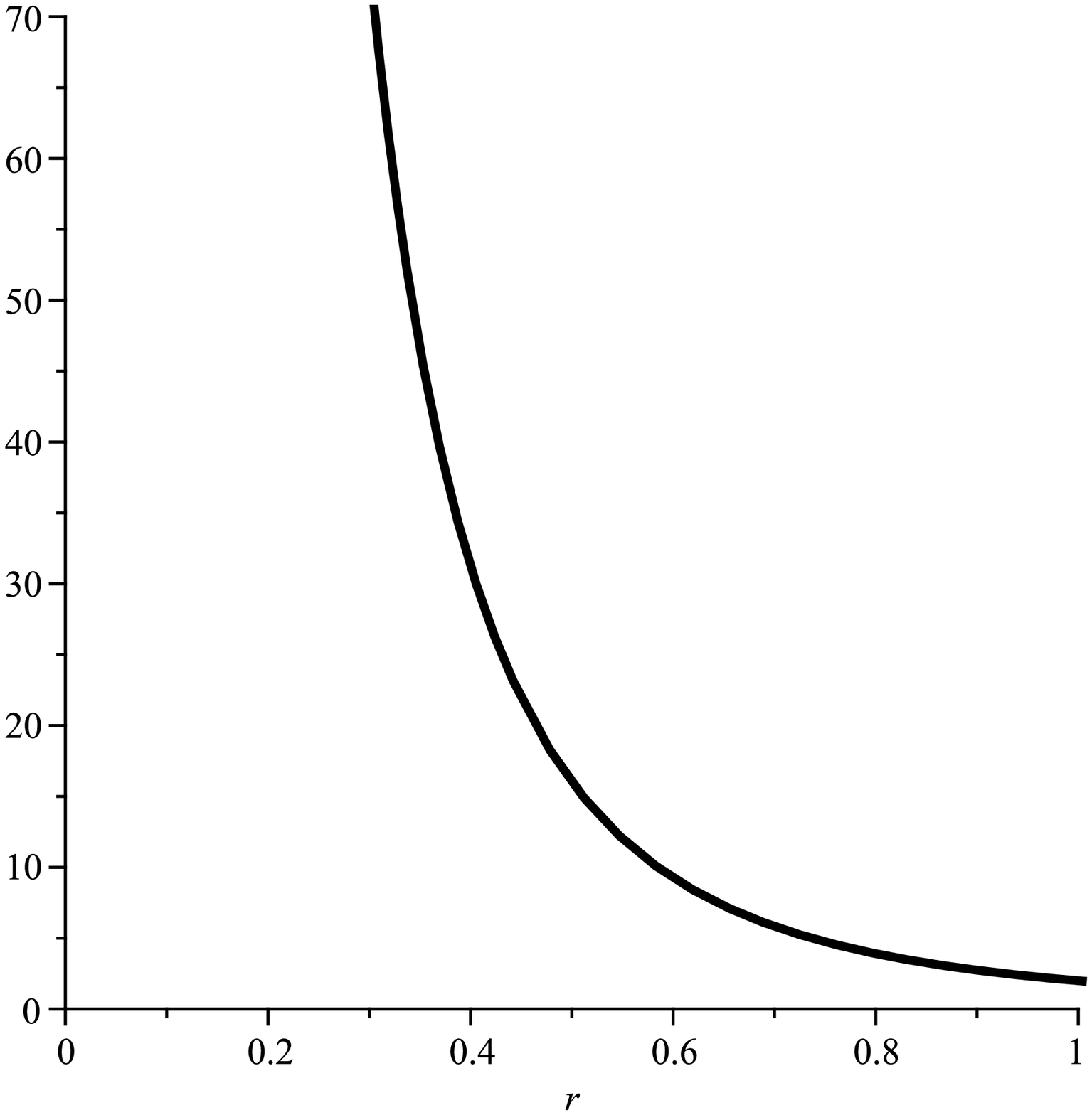}
\hskip.3cm
\includegraphics[scale=0.3]{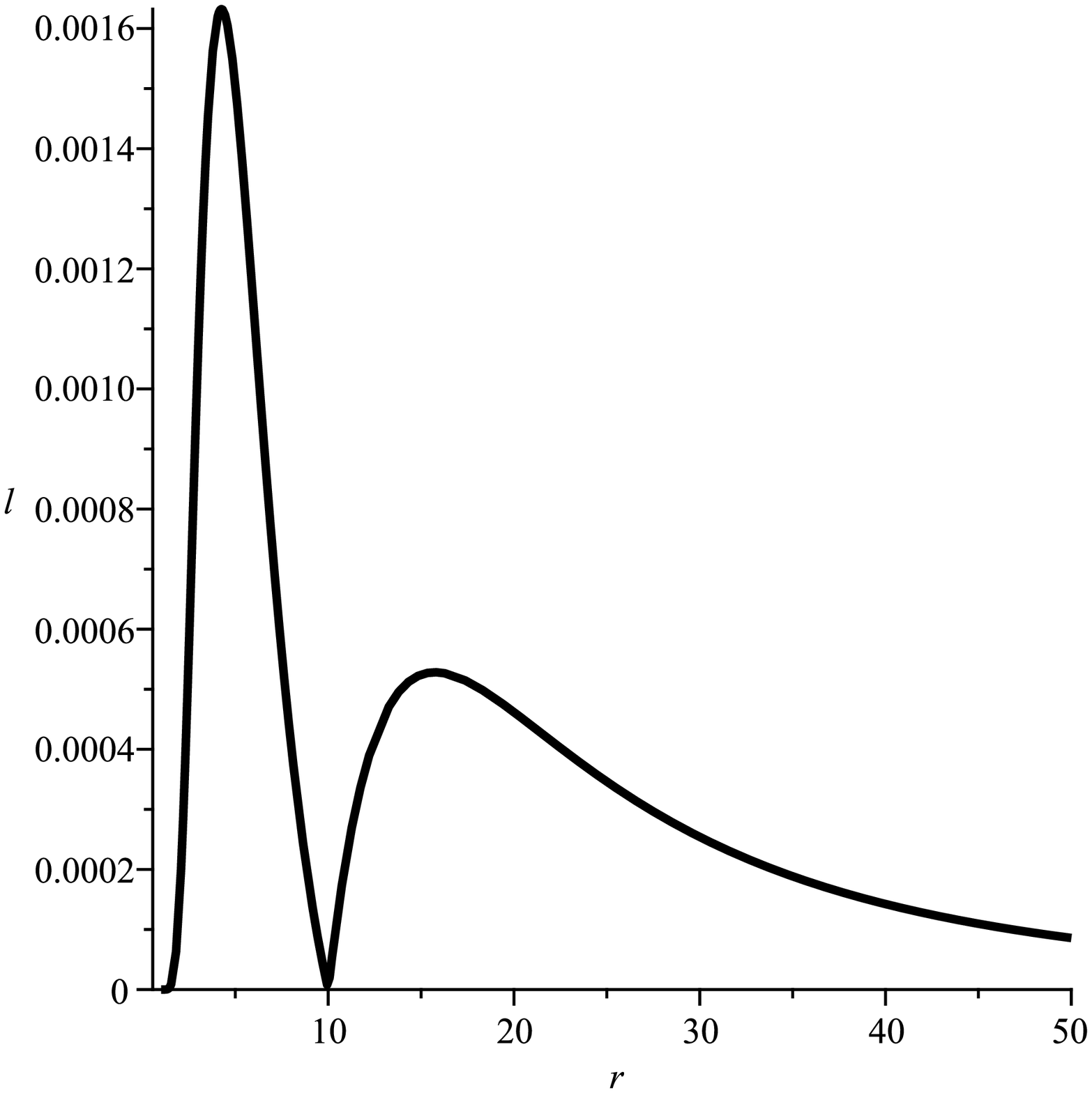}
\caption{Behavior of the curvature eigenvalue on the equatorial plane $(\theta=\pi/2)$
 of the Zipoy-Voorhees metric. Plot (a) corresponds to a
black hole solution with $q=0$. Plot (b) illustrates the behavior in case of a naked singularity with $q=-2$.}
\label{fig3}
\end{figure}

Whereas the effective potential of a black corresponds
to the typical potential of an attractive field, the effective potential of a naked singularity
is characterized by the presence of a barrier which acts on test particles as a source of repulsive gravity. 
Although this result is very intuitive, the disadvantage of this analysis is that it is not invariant. In fact, a coordinate transformation can be used to arbitrarily change the position of the barrier of repulsive gravity. Moreover, the identification
of the spatial coordinate $r$ as a radial coordinate presents certain problems in the case of metrics with quadrupole moments
\cite{novzel}. 
To avoid this problem we investigate a set of scalars that can be constructed from the curvature tensor and are linear in the 
parameters that enter the metric, namely, the eigenvalues of the Riemann tensor. Let us recall that the curvature of the Zipoy--Voorhees metric belongs
to type I in Petrov's classification. On the other hand, type I metrics  possess three different curvature eigenvalues whose real parts are scalars \cite{jordan}. The explicit calculation of the curvature eigenvalues for this metric 
shows \cite{quev90} that all of them are real
and, consequently, they behave as scalars under arbitrary diffeomorphisms.  The resulting
analytic expressions are rather cumbersome. For this reason we performed a numerical analysis and found out the main differences
between black holes and naked singularities.
The results are illustrated in Fig. \ref{fig3}.

We took a particular eigenvalue which represents the qualitative behavior of all 
the eigenvalues. In the case of a black hole, the eigenvalue diverges near the origin of coordinates, where the curvature singularity is situated, and it decreases rapidly as $r$ increases, tending to zero at spatial infinity. 
In the case of a naked singularity the situation changes drastically. The eigenvalue vanishes at spatial infinity and then increases as the value of the radial coordinate decreases. At a specific radius $r=r_{min}$, the eigenvalue reaches a local maximum and then 
rapidly decreases until it vanishes. This oscillatory behavior becomes more frequent as the origin of coordinates is approached. 
It seems plausible to interpret this peculiar behavior as an invariant manifestation of the presence of repulsive gravity.
On the other hand, if one would like to avoid the effects of repulsive gravity, one would propose $r_{min}$ as the 
minimum radius where the matching with an interior solution should be carried out. If we denote the eigenvalue as $\lambda$,
then $r_{min}$ can be defined invariantly by means of the equation
\be
\frac{\partial \lambda}{\partial r}\bigg| _{r=r_{min}}=0\ .
\label{matcon}
\ee
Then, the radius $r_{min}$ determines the matching hypersurface $\Sigma$ and one could interpret condition (\ref{matcon})
as a $C^3-$matching condition. In concrete cases, one must calculate all possible eigenvalues $\lambda_i$ and all possible 
points satisfying the matching condition ${\partial \lambda_i}/{\partial r}=0$. The radius $r_{min}$ corresponds then to the first 
extremum that can be found when approaching the origin of coordinates from infinity.  In the next section we will show that 
this approach can be successfully carried out in the case of the Zipoy--Voorhees metric.

\subsection{An interior solution}
\label{sec:interior}

In the search for an interior solution that could be matched to the exterior solution with quadrupole moment given in 
Eq.(\ref{zv}),  we found that an appropriate
form of the line element can be written as
\be
ds^2 = fdt^2 - \frac{e^{2\gamma_0}}{f}\left(\frac{dr^2}{h} + d\theta^2\right) -\frac{\mu^2}{f}d\varphi^2\ ,
\ee
where 
\be
e^{2\gamma_0} = (r^2-2mr+m^2\cos^2\theta)e^{2\gamma(r,\theta)}\ ,
\ee
and $f=f(r,\theta)$, $h=h(r)$, and $\mu=\mu(r,\theta)$. This line element preserves axial symmetry and 
staticity. 

The inner structure of the mass distribution  
with a quadrupole moment can be described by a perfect fluid energy--momentum tensor (\ref{pf}). In general, 
in order to solve Einstein's equations completely, 
pressure and energy must be functions of the coordinates $r$ and $\theta$. 
However, if we assume that 
$\rho=$ const, the resulting system of differential equations is still compatible. 
The assumption of constant density drastically reduces the complexity of the problem. 
Then, the corresponding field equations reduce to
\be
p_r = - \frac{1}{2} (p+\rho) \frac{f_r}{f}\ , \quad p_\theta =  - \frac{1}{2} (p+\rho) \frac{f_\theta}{f}\ ,
\ee
\be
\mu_{rr} = -\frac{1}{2h} \left( 2 \mu_{\theta\theta} + h_{r} \mu_r - 32 \pi p\frac{\mu e^{2\gamma_0}}{f} \right) \ ,
\ee
\be
f_{rr} = \frac{f_r^2}{f} -\left(\frac{h_r}{2h} + \frac{\mu_r}{\mu}\right)f_r + \frac{f_\theta^2}{hf} -
\frac{\mu_\theta f_\theta}{\mu h} -\frac{f_{\theta\theta}}{h} + 8\pi \frac{(3p+\rho)e^{2\gamma_0}}{h}\ .
\ee
Moreover, the function $\gamma$ turns out to be determined by a set of two partial differential equations
which can be integrated by quadratures once $f$ and $\mu$ are known. The integrability condition of these partial
differential equations turns out to be satisfied identically by virtue of the remaining field equations.

\begin{figure}
\begin{center}
\includegraphics[scale=0.4]{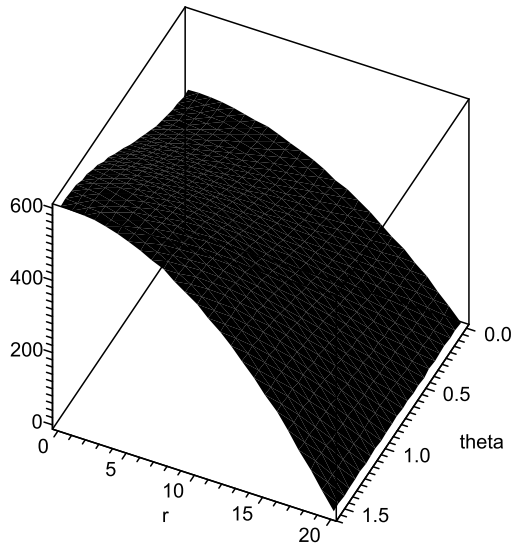}
\end{center}
\caption{Plot of the inner pressure as a function of the spatial coordinates.
\label{fig4}
}
\end{figure}

Although we have imposed several physical conditions which simplify the form of the field equations, we 
were unable to find analytic solutions. However, it is possible to perform a numerical integration by imposing
appropriate initial conditions. In particular, we demand that the metric functions and the pressure are finite 
at the axis. 
Then, it is possible to plot all the metric functions and thermodynamic variables. 
In particular, the pressure behaves as shown 
in Fig.\ref{fig4}.

It can be seen that the pressure is finite in the entire interior domain, and tends to zero at certain hypersurface $R(r,\theta)$ 
which depends on the initial value of the pressure on the axis. Incidentally, it turns out that by increasing the value of the 
pressure on the axis, the ``radius fuction" $R(r,\theta)$ can be reduced. Furthermore, if we demand that the hypersurface $R(r,\theta)$ 
coincides with the origin of coordinates, the value of the pressure at that point diverges. 
From a physical point of view, this is exactly the behavior that 
is expected from a physically meaningful  pressure function. 

\begin{figure}
\begin{center}
\includegraphics[scale=0.3]{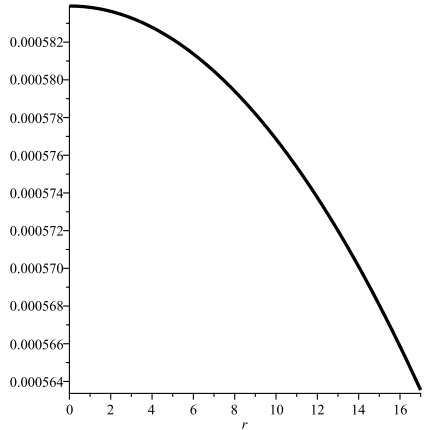}
\end{center}
\caption{Behavior of the curvature eigenvalue on the equatorial plane $(\theta=\pi/2)$
 of the interior solution.
 \label{fig5}
 } 
\end{figure}

This solution can be used to calculate numerically the corresponding Riemann tensor and its eigenvalues. As a result we obtain
that the solution is free of singularities in the entire region contained within the radius function $R(r,\theta)$. In particular,
one of the eigenvalues presents on the equatorial plane the behavior depicted in Fig.\ref{fig5}. All the eigenvalues have a finite value at the symmetry axis and decrease as the boundary surface is approached.

\begin{figure}
\begin{center}
\includegraphics[scale=0.3]{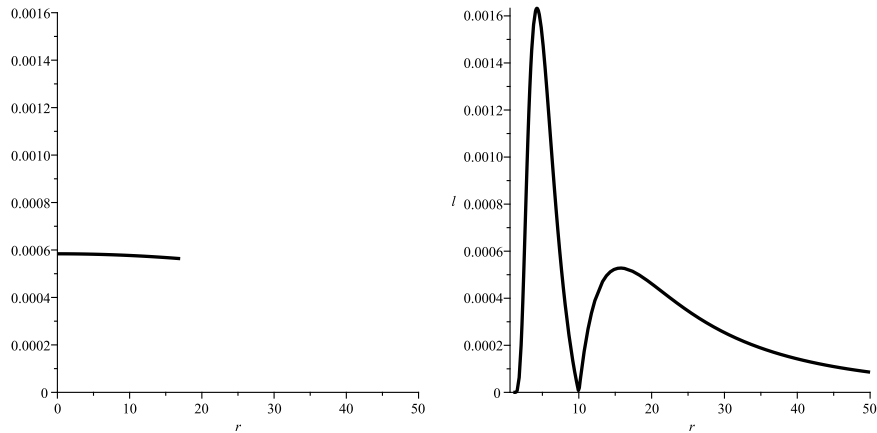}
\includegraphics[scale=0.3]{zvinv10.eps}
\end{center}
\caption{Curvature eigenvalues of the interior solution and of the exterior solution with the same scale.
\label{fig6}
}
\end{figure}

To apply the $C^3-$matching procedure proposed above we compare the behavior of the eigenvalue plotted in Fig.\ref{fig3} with
the corresponding eigenvalue plotted in Fig.\ref{fig5}, using the same scale in both graphics. The result is illustrated in Fig.\ref{fig6}. It then becomes clear that the first possible point where the matching can be performed is exactly at $r_{min}$ 
which in this particular case corresponds to $r_{min} \approx 5M_0$. This fixes the initial value of the pressure on the axis which is then used 
to attack the problem of matching the interior and exterior metric functions. In all the cases we analyzed, we obtained
 a reasonable matching, withing 
the accuracy of the numerical calculations. We repeated the same procedure for different values of the angular coordinate  ($\theta=\pi/4$ and 
$\theta =0$), and obtained 
that the matching can always be reached by fixing in an appropriate manner the arbitrary constants that enter the metric 
functions $f$ and $\mu$.

\section{Conclusions}
\label{sec:con}

In this work, we presented an exact electrovacuum solution of Einstein-Maxwell equations which 
contains four different sets of multipole moments. An invariant calculation shows that they
can be interpreted as the gravitoelectric, gravitomagnetic, electric and magnetic multipole
moments. The solution is asymptotically flat and is free of singularities in a region situated
around the origin of coordinates. The rotating Kerr metric is contained as a special case. 
The NUT parameter can also be included by a suitable choice of the arbitrary constants 
which enter the Ernst potentials. We conclude that this solution can be used to 
describe the exterior gravitational field a charged rotating mass distribution. 

In the particular case of slowly rotating and slightly deformed mass distribution we obtained
the explicit form of the metric, and showed that it can be matched with an interior solution
which is contained within the class of Hartle-Thorne solutions. This reinforces the conclusion
that the solution represents the interior as well as the exterior 
gravitational field of astrophysical compact objects.

We study the problem of matching the interior and exterior spacetimes. We propose a $C^3-$matching
which consists in demanding that the derivatives of a particular curvature eigenvalue are smooth
on the matching hypersurface. To prove the validity of this approach we derived an interior solution
for the simplest case of a static mass with an arbitrary quadrupole moment, represented by the
Zipoy--Voorhees vacuum solution. The numerical integration of the corresponding field equations shows 
that interior perfect fluid solutions exist which are characterized by a constant density profile with a
variable pressure. Fixing the value of the angular coordinate $\theta$, we performed numerically 
the $C^3-$matching. As a result we obtain a minimum radius at which the matching can be carried out and 
a fixed value for the pressure on the symmetry axis. These values are then used to reach the smooth matching 
of the interior and exterior metric functions. In all the cases analyzed in this manner we obtained a reasonable 
numerical matching. 

The idea of using the $C^3-$matching condition to determine the minimum radius, at which an interior solution
can be matched with an exterior one, has been proved also in a particular case where analytical methods
can be applied, namely, in the case of the Kerr-Newman class of solutions. The obtained results are reasonable 
and compatible with other results obtained by analyzing the motion of test particles \cite{luoquev10}. 
These results indicate that it should be possible to determine the minimum radius of an astrophysical 
compact object by using the idea of the $C^3-$matching presented here. To prove this conjecture 
in general, it will be necessary to use more powerful  methods related to the mathematical 
behavior of geodesics and curvature. This problem is currently under investigation \cite{kqr10}. An 
important application of this analysis would be  to relate the minimum size of a compact object with 
its binding energy. As a result we would obtain the maximum binding energy which is physically allowed
for an astrophysical compact object.

\section*{Acknowledgments}
I would like to thank D. Bini, A. Geralico, R. Kerr, O. Luongo, and R. Ruffini for helpful comments.
I also thank ICRANet for support.




\begin{thebibliography}{99}

\bibitem{solutions}
H. Stephani, D. Kramer, M. A. H. MacCallum, C. Hoenselaers, and E. Herlt, 
{\it Exact Solutions of Einstein's Field Equations} (Cambridge University Press, Cambridge, UK, 2003).

\bibitem{qm91}
H. Quevedo and B. Mashhoon, {\it Generalization of Kerr spacetime},  Phys. Rev. D {\bf 43}, 3902 (1991).

\bibitem{ernst} F. J. Ernst, {\it New formulation of the axially symmetric 
gravitational field problem}, {\it Phys. Rev.} {\bf 167} (1968) 1175;
{\it New Formulation of the axially symmetric gravitational field problem II},
Phys. Rev. {\bf 168} (1968) 1415.



\bibitem{zip66} D. M. Zipoy, 
{\it Topology of some spheroidal metrics}, 
J. Math. Phys. {\bf 7}, 1137 (1966).

\bibitem{voor70} B. Voorhees, 
{\it Static axially symmetric gravitational fields}, 
Phys. Rev. D {\bf 2}, 2119 (1970).

\bibitem{erro59} G. Erez and N. Rosen, Bull. Res. Counc. Israel {\bf 8}, 47 (1959).



\bibitem{ger} R. Geroch, 
{\it Multipole moments. I. Flat space}
{J.\ Math.\ Phys.} {\bf 11}, 1955 (1970);
{\it Multipole moments. II. Curved space}, {J.\ Math.\ Phys.} {\bf 11}, 2580 (1970).

\bibitem{hans} R. O. Hansen, {\it Multipole moments of stationary spacetimes}
{J.\ Math.\ Phys.} {\bf 15}, 46 (1974).



\bibitem{quev89} H. Quevedo, {\it General Static Axisymmetric Solution of Einstein's Vacuum
Field Equations in Prolate Spheroidal Coordinates}, Phys. Rev. D {\bf 39}, 2904--2911 (1989).

\bibitem{bglq09} D. Bini, A. Geralico, O. Luongo, and H. Quevedo, {\it Generalized Kerr spacetime with an arbitrary quadrupole moment:
Geometric properties vs particle motion}, Class. Quantum Grav. {\bf 26}, 225006 (2009).

\bibitem{quev10} H. Quevedo, {\it Exterior and interior metrics with quadrupole moment}, Gen. Rel. Grav. (2010), in press.

\bibitem{hartle1} J. B. Hartle, {\it Slowly rotating relativistic stars:  I. Equations of structure},  
Astrophys. J. {\bf 150}, 1005 (1967).

\bibitem{hartle2} J. B. Hartle and K. S. Thorne, 
{\it Slowly rotating relativistic stars:  II. Models for neutron stars and supermassive stars},
Astrophys. J. {\bf 153}, 807 (1968).



\bibitem{joshi07} P. S. Joshi, {\it Gravitational Collapse and Spacetime Singularities} 
(Cambridge University Press, Cambridge, 2007).


\bibitem{def89} F. de Felice, {\it Repulsive gravity and curvature invariants in general relativity}, Ann. de Phys. {\bf 14}, 79 (1989).


\bibitem{def08} G. Preti and F. de Felice, {\it Light cones and repulsive gravity}, Am. J. Phys. {\bf 76}, 671 (2008).


\bibitem{christian} C. Cherubini , D. Bini,  S. Capozziello,  and R. Ruffini,
{\it Second order scalar invariants of the Riemann tensor: Applications to black hole space-times},
Int. J. Mod. Phys. D {\bf 11}, 827 (2002). 

\bibitem{novzel} Ya. V. Zeldovich and I. D. Novikov, {\it Relativistic Astrophysics} (University of Chicago Press, Chicago,
1971). 

\bibitem{jordan} P. Jordan, J. Ehlers and W. Kundt, Akad. Wiss. Main. Abh. Math.-Nat. (1960); see also Ref.[1]. 

\bibitem{quev90} H. Quevedo, {\it Multipole moments in general relativity - Static and stationary solutions -}, Forts. Phys.  {\bf 38}, 733 (1990).

\bibitem{luoquev10} O. Luongo and H. Quevedo, {\it Toward an invariant definition of repulsive gravity}, in Proceedings
of the 12-th Marcel Grossman Meeting on General Relativity and Gravitation (2010). 

\bibitem{kqr10} R. P. Kerr, H. Quevedo, and R. Ruffini, {\it On the minimum size of astrophysical compact objects},
in preparation

\end{thebibliography}
\end{document}